# ORAN-Bench-13K: An Open Source Benchmark for Assessing LLMs in Open Radio Access Networks

Pranshav Gajjar and Vijay K. Shah
*NextG Wireless Lab*, George Mason University
pgajjar@gmu.edu and vshah22@gmu.edu

*Abstract*—Large Language Models (LLMs) can revolutionize how we deploy and operate Open Radio Access Networks (O-RAN) by enhancing network analytics, anomaly detection, and code generation and significantly increasing the efficiency and reliability of a plethora of O-RAN tasks. In this paper, we present ORAN-Bench-13K, the first comprehensive benchmark designed to evaluate the performance of Large Language Models (LLMs) within the context of O-RAN. Our benchmark consists of 13,952 meticulously curated multiple-choice questions generated from 116 O-RAN specification documents. We leverage a novel three-stage LLM framework, and the questions are categorized into three distinct difficulties to cover a wide spectrum of ORAN-related knowledge. We thoroughly evaluate the performance of several state-of-the-art LLMs, including Gemini, Chat-GPT, and Mistral. Additionally, we propose ORANSight, a Retrieval-Augmented Generation (RAG)-based pipeline that demonstrates superior performance on ORAN-Bench-13K compared to other tested closed-source models. Our findings indicate that current popular LLM models are not proficient in O-RAN, highlighting the need for specialized models. We observed a noticeable performance improvement when incorporating the RAG-based ORANSight pipeline, with a Macro Accuracy of 0.784 and a Weighted Accuracy of 0.776, which was on average 21.55% and 22.59% better than the other tested LLMs.

## I. INTRODUCTION

The telecommunications landscape is undergoing a paradigm shift with the emergence of Open Radio Access Networks (ORAN), a disruptive approach promoting openness, flexibility, and innovation in mobile network architectures [1]. ORAN's modular and interoperable framework enables operators to integrate components from diverse vendors, fostering a more dynamic and cost-effective ecosystem. As O-RAN gains traction globally, its significance in shaping the future of mobile communications cannot be overstated [1] [2]. Concurrently, the advent of Large Language Models (LLMs) has sparked a revolution in Natural Language Processing (NLP) and Artificial Intelligence (AI), elevating text generation, comprehension, and interaction to unprecedented levels of sophistication [3]. Large Language Models (LLMs), such as OpenAI's Generative Pre-trained Transformer (GPT) series and open-source variants like the Mistral family of models, have demonstrated remarkable capabilities in understanding and generating human-like text [4] [5].

We have seen the impact of LLMs and related NLP technologies in multiple domains like Finance and medicine [6] with revolutionary applications and a recent advent in research for the telecommunications and wireless industry [6]. LLMs have been used for automated code refactoring and design [7], recommending troubleshooting solutions [8], generating network configurations [9], optimization tasks like load balancing [10], and even for prediction-based beamforming [11], and traffic load prediction [12]. The domain of O-RAN is still in the nascent stages of LLM applications with papers working towards Intent Processing and Network Optimization [13], and countless surveys on the possible avenues of research for LLMs in O-RAN and related avenues [14] [13] [15].

It is important to understand that the inclusion of such LLMs is only possible after creating thorough evaluation strategies as existing benchmark datasets in multiple domains have played a crucial role in assessing model performance and guiding architectural design [6], with prominent examples being FLUE [16] and MultiMedQA [17]. Evaluating performance for these specialized language models is not as straightforward as traditional ML applications [18], and It is extremely important to have a comprehensive evaluation strategy to deploy such LLMs due to the excessive computational costs that are observed during retraining and architecture creation. It is possible to assess the efficacy of LLM-based solutions by qualitatively observing the generated outputs by human feedback. However, the process would require immense effort and resources to analyze comprehensively and still be susceptible to human errors and biases. Due to the hallucination problem [19] in LLMs, without skilled evaluators, assessing an LLM's performance for a specific domain would become difficult, and inaccurate. There are a few papers that have leveraged an LLM-based or a generative solution to create Multiple Choice Question benchmarks by using a large corpus of documents, with promising results. The use of the MCQ styles benchmark has also been heavily proposed in the literature [17] [6].

Considering the impact LLMs can have in O-RAN and due to the absence of any evaluation or benchmarking tools, this paper offers the following primary contributions:
- It aims to create a novel benchmark named ORAN-Bench-13K that can accurately assess an LLM's performance for O-RAN specifications knowledge. This is constructed by leveraging three different LLM instances resulting in a total of **13952** MCQs spanning three different difficulty levels.
- The paper also assesses different publicly available and closed-source LLMs, that have been proven useful in a multitude of applications including Chat-GPT, Gemini-1.5, and Mistral-7B.
- We also propose a novel Retrieval Augmented Generation (RAG) [20] based LLM pipeline called ORANSight that operates on a Facebook AI Similarity Search (FAISS) database [21] and a corpus of specifications documents to understand the efficacy of RAG for O-RAN and the possibility of creating an O-RAN Centric LLM.
- Our findings entail that though models like Chat-GPT and Gemini-1.5 [22] are not proficient enough in O-RAN knowledge, we could observe state-of-the-art (SOTA) performance from our ORANSight solution.

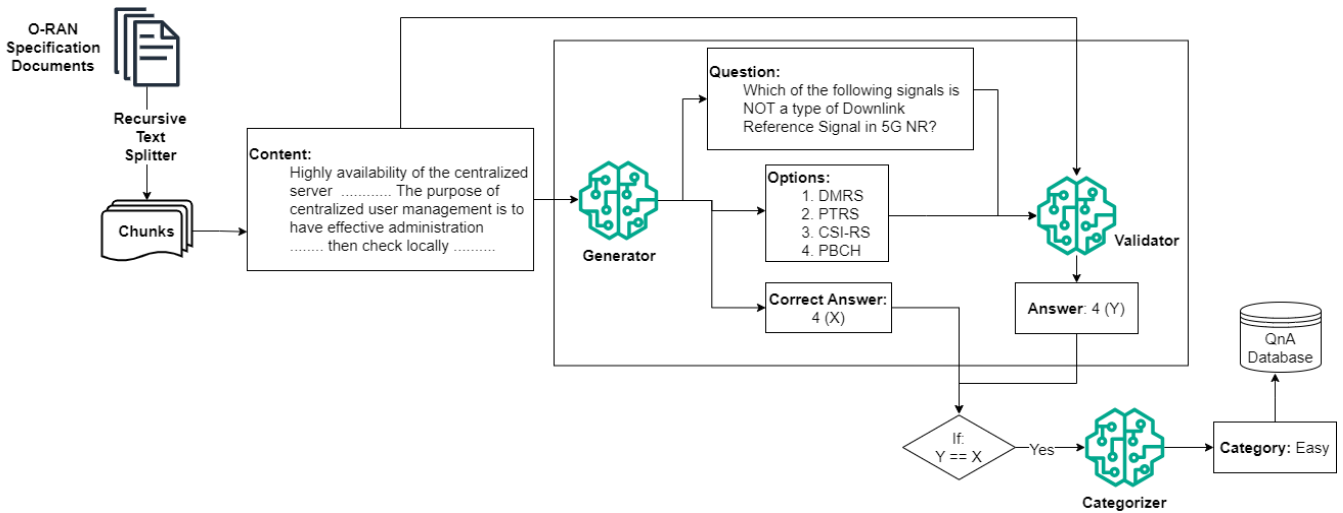

Fig. 1. A high-level overview for the proposed MCQ Generation Pipeline.

Staying abreast of the latest O-RAN specifications is also both crucial and challenging. The sheer volume and complexity of these documents can be overwhelming, making it difficult for engineers, researchers, and industry professionals to efficiently extract and apply the necessary information [23]. The use of LLMs that are well-versed in O-RAN specifications can offer significant benefits and assist in rapid development. As the incorporation of RAG with LLMs would help in inducing domain knowledge, we also explore the possibility of leveraging ORANSight for a Specification Assistant as an auxiliary task.

**Paper Organization.** This paper is organized as follows: Section II presents the Data Sources, Section III provides an overview of the Benchmark Generation process, and Section IV discusses the proposed RAG framework, ORANSight. Other tested LLMs are outlined in Section V, followed by the presentation of results in Section VI. Concluding remarks are offered in Section VII, and sample questions from the benchmark are included in Section VIII.

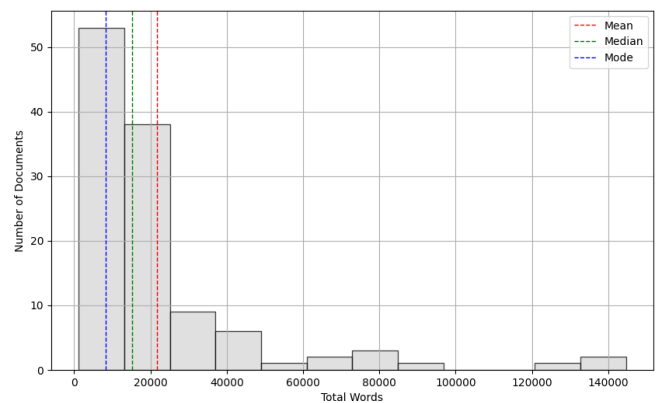

Fig. 2. Central Tendencies for the specification documents and total words.

## II. DATA SOURCES

We source all available specification documents from [24] resulting in a total of 116 documents with an average of 21778 words per specification document and a total of 2.53 million words. The data processing pipeline will be explained thoroughly in the subsequent sections, and the central tendencies of the total number of words are mentioned in Figure 2.

## III. BENCHMARK GENERATION

The MCQ generation is performed in a multi-stage process as depicted in the Figure 1. The primary aspect of this pipeline is inspired by the papers [23] [6] and contains the generator and validator pair of LLMs which would determine the total number of acceptable generated questions. All the LLM instances in the generation process use Gemini-1.5 as it has a higher yield rate than a Mistral-7B from initial experiments. The entire process can be explained as:
- The specification documents are converted to small chunks of text, with a size of 1536 and an overlap of 256 characters. This specific number for obtained by experimenting with different context sizes, as a larger number would result in losing out on important O-RAN content, and a smaller value would simply not have enough meaning to generate a valid MCQ. The overlap is added to preserve semantic context between chunks.
- **Generator LLM**: This is the first LLM instance in this system, which is prompted to use the provided content or chunk to generate a valid MCQ specific for an O-RAN benchmark. It is only prompted to generate one question per chunk to avoid repetition. The generated text is then parsed for the Question, Options, and Answer tags,
- **Validator LLM**: This is solely responsible for assessing the generated questions, it is provided the context, the generated questions, and the options, and prompted to generate an answer as an Extractive QnA [25] task. This shows us if the generated question has semantic meaning with the provided chunk, and if the options and answer are coherent as well. If the validator's answer and the generated answer are the same we proceed with the next steps, or else the question is rejected.
- **Categorizer LLM**: This model is prompted to segregate the valid questions into three categories based on the descriptions:
  - *Easy*: Questions that focus on basic concepts and known facts.
  - *Intermediate*: Questions that require comprehension and application of concepts, or involve moderate calculations or reasoning.
  - *Difficult*: Questions that demand a deep understanding of Open RAN standards, or the ability to synthesize multiple pieces of information.
- The final outputs are appended to a database and the process on the aforementioned O-RAN Specification documents results in **1139** 'Easy', **9570** 'Intermediate', and **3243** 'Difficult' questions.

The entire benchmark is available online at [1] along with the required supplementary files.

## IV. ORANSIGHT

The proposed RAG framework consists of three main parts, the Embedding Generator, the FAISS database, and the Mistral-7B LLM. The reason we leverage the Mistral model is to keep ORANSight open source and easily adaptable to existing LLM-based O-RANuse cases. We use a **BGE**-Small-1.5 (BAAI General Embeddings) model [26] as our embedding generator, BGE stands for three model sizes: small (24M), base (102M), and large (326M), representing an embedding dimension of 384, 786, and 1024 respectively [26]. We choose BGE to main an open-source implementation and the small variant for the increased computational efficiency.

**FAISS** can be perceived as an efficient library for fast similarity search and clustering of dense vectors. As it is designed to handle large-scale data, FAISS is optimized for both memory usage and computational speed, making it ideal for high-dimensional text embeddings [21]. The core functionality of FAISS includes indexing methods that enable rapid approximate nearest neighbor (ANN) searches, allowing for real-time retrieval of relevant vectors that can be used to obtain the original text which is a document chunk. FAISS has been used as a critical component for various RAG-based systems [21] further validating its use in ORANSight.

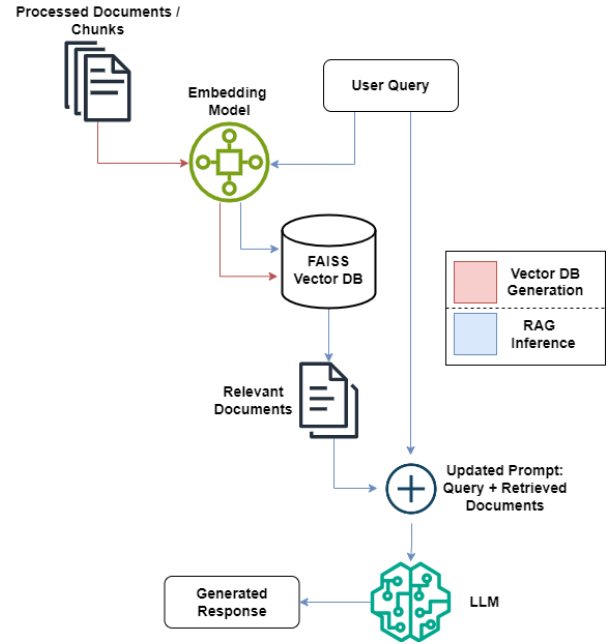

Fig. 3. A high-level overview for ORANSight inference and database generation.

The inherent function of ORANSight as depicted in the Figure 3 can be explained as:
- **Embedding Generation**: Here all document chunks are also obtained by a Recursive Text Splitter (using a chunk size of 1024, and an overlap of 256) and are converted into a set of 1D vectors using a BGE instance with a dimension of 384.
- **Inference**: Once a processed database is ready, for each user query, the top 5 relevant documents are retrieved and appended into the query resulting in a new prompt. The final prompt with added context is used by the Mistral-7B to make predictions.

To implement a conversational chatbot through the RAG computational chain, we append the system with a Buffer Memory [27] which would facilitate extended conversations and initial tests on ORANSight as a specification assistant.

## V. TESTED LLMs

We experiment with three main models, ChatGPT-3.5, Gemini-1.5, and Mistral-7B, all three are instruction-tuned. The total parameter and architectural specifications aren't available but the utility of such general-purpose models can be seen in a variety of domains including telecommunications. The Mistral-7B on the other hand is also based on the Transformer architecture and leverages Sliding Window Attention, which is considerably faster to compute than the vanilla variant [5]. It also uses a Rolling Buffer Cache which further enhances the computational efficiency while offering SOTA results on general-purpose NLP benchmarks.

---
[1]https://github.com/prnshv/ORAN-Bench-13K

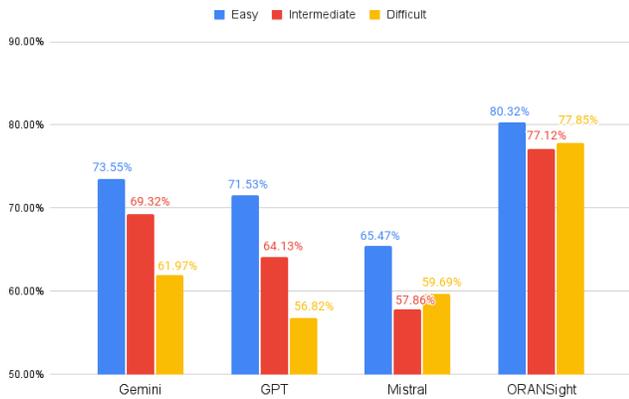

Fig. 4. Primary Accuracy Scores, here, ORANSight indicates a Mistral + RAG model.

## VI. RESULTS

This section elaborates on the obtained results using different LLMs on our benchmark. All the experiments were conducted by leveraging the Langchain library [27], GPT was implemented using the OpenAI extension, Gemini by the Google-GenAI extension, and both BGE and Mistral-7B were implemented using Huggingface [28].

As shown in Figure 4, it can be inferred that ORANSight (i.e., using RAG with Mistral) is the best-performing model with superlative results in all three categories. There was an absolute increase of 14.85%, 19.26 %, and 18.17% across Easy, Intermediate and Difficult categories. It is also important to note that a Mistral model only has 7 billion parameters, when compared with other prominent architectures which are estimated to have at least 175 billion parameters [29], the results hold immense value. The Gemini model was obtained to be better suited for O-RAN than ChatGPT as it was relatively better across all three categories. The worst-performing model was a vanilla Mistral and both the baseline and the RAG-enhanced model performed relatively better on the difficult benchmark than the intermediate questions.

TABLE I
CUMULATIVE PERFORMANCE OF ALL TESTED MODELS.

| Model | Macro Accuracy | Weighted Accuracy |
|---|---|---|
| Gemini | 0.683 | 0.680 |
| GPT | 0.642 | 0.630 |
| Mistral | 0.610 | 0.589 |
| **ORANSight** | **0.784** | **0.776** |

The results were further reinforced by the table I which showcases the Macro and Weighted scores for all tested models. The weighted accuracy is calculated by using the no.of generated questions as weight values which can be referred to in section III. The Macro Accuracy symbolizes an average score without assuming any weight values. The ORANSight pipeline had an average increase in performance by the magnitude of 21.55% and 22.59% for Macro and Weighted scores. These findings underscore the effectiveness of the RAG-based approach in enhancing LLM performance within the specialized domain of ORAN. While the evaluated models, including Gemini, GPT, and Mistral, were better than a random classifier (signifying a 25% score) it is still far from optimal. This further highlights the necessity for developing and utilizing specialized models to achieve higher accuracy in domain-specific tasks within the O-RAN context.

## VII. CONCLUSIONS AND FUTURE WORK

This paper aimed to create a comprehensive benchmark for evaluating the performance of LLMs for ORAN-centric tasks. By leveraging a novel multi-stage process with three LLM instances we have created a benchmarking set consisting of 13952 Multilpe Choice Questions. These were further segregated into three distinct difficulty criteria, Easy, Intermediate, and Difficult symbolizing varying degrees of proficiency with 1139, 9570, and 3243 questions respectively. We also propose a RAG-based pipeline named ORANSight that would add domain-specific knowledge to a Mistral-7B model. After a thorough analysis of three different LLM architectures Chat-GPT, Gemini-1.5, and Mistral 7B, we conclude that though the benchmark scores are better than a random classifier, the LLMs are not proficient in O-RAN. A significant performance boost can be obtained by leveraging RAG as we could observe an average increase of 21.55% and 22.59% for both macro and weighted scores against the other tested models. For future work, we wish to create an open-source fine-tuning dataset that can be used to train O-RAN-proficient LLMs and also explore a coding-based benchmark that can assess a model's ability to perceive O-RAN codes.

## VIII. ACKNOWLEDGEMENT

The project is partially supported by the Public Wireless Supply Chain Innovation Fund under Federal Award ID Number 26-60-IF010.

## IX. APPENDIX: A

This section contains sample questions from the ORAN-Bench-13K for all the aforementioned categories.

### A. Easy

**Question**: In the context of O-RAN, what is the primary purpose of the Alarm List?
**Options**:
1) To store a history of alarm events detected by the IMS.
2) To manage the configuration of logging levels for various O-RAN components.
3) To provide a centralized repository for all O-RAN network performance metrics.
4) To facilitate the real-time monitoring of network traffic patterns.

**Answer**: 1

**Question**: Which of the following protocols is used for communication between the O-DU and O-RU in an O-RAN network?

**Options**:

1) F1AP
2) NGAP
3) FH-eCPRI
4) HTTP2

**Answer**: 3

**Question**: Which component of the O-RAN architecture is responsible for controlling the radio access network in near real-time?

**Options**:

1) gNB-CU
2) Near-RT RIC
3) O-CU-CP
4) FHGW

**Answer**: 2

*B. Intermediate*

**Question**: Which of the following is a method used by the Service Management and Orchestration Framework to dynamically operate and maintain an O-RAN network?

**Options**:

1) Configuring IP addressing for PNFs and VNFs.
2) Managing software updates for existing NFs.
3) Adding, removing, or modifying NFs.
4) All of the above.

**Answer**: 4

**Question**: What is the purpose of the NESPolicy Information Object Class (IOC) in O-RAN?

**Options**:

1) To configure the callHomeClientInfo data type.
2) To provide policies for enabling or disabling energy-saving features in the O-RU.
3) To manage the Shared O-RU Host role configuration.
4) To define attributes for the Shared O-RU Host role.

**Answer**: 2

**Question**: In a ladder topology for Open Radio Access Networks (O-RAN), what is the primary purpose of having redundant sync flows from two different GMs?

**Options**:

1) To improve the overall capacity of the network.
2) To ensure failover and extended sync outages.
3) To enable carrier aggregation across different leaf networks.
4) To reduce latency in data transmission.

**Answer**: 2

*C. Difficult*

**Question**: What does the "L_USE" status of the sync-source-status parameter in the ptp-status container indicate?

**Options**:

1) The device is synchronized to an alternate master clock.
2) The device is synchronized to the current master clock.
3) The device is not synchronized to any master clock.
4) The device is attempting to synchronize to an alternate master clock.

**Answer**: 2

**Question**: Which of the following is the purpose of the OR.CellUA.MinPscellUeUl performance counter in O-RAN?

**Options**:

1) To measure the maximum number of Primary Serving Cells (PSCells) in a cell.
2) To track the minimum number of User Equipment (UEs) connected to a PSCell in the uplink (UL) every 100 milliseconds.
3) To monitor the average number of UEs connected to a PSCell over a specified time period.
4) To calculate the total number of UEs in a cell.

**Answer**: 2

**Question**: Which of the following is NOT a required testing tool for verifying the removal of an SN terminated split bearer initiated by the MN (eNB)?

**Options**:

1) Test UE or UE emulator capable of supporting both LTE and NR
2) Core or Core emulator (EPC with EN-DC capabilities)
3) Application test server for UDP and TWAMP traffic
4) Radio Frequency Analyzer for observing RF signals

**Answer**: 4

X. APPENDIC: B

This section contains sample text generated by ORANSight indicating a possible use case as a specification assistant.

**Prompt:** Could you clarify the differences between the O-RAN fronthaul and midhaul interfaces?

**Generated Response:** The O-RAN fronthaul interface is the connection between the O-DU (Distributed Unit) and the O-RU (Remote Unit) in the O-RAN (Open Radio Access Network) architecture. It is responsible for transporting user data and control information between the O-RU and the O-DU.

On the other hand, the O-RAN midhaul interface is the connection between the O-CU (Central Unit) and the O-DU. It is responsible for transporting user data and control information between the O-CU and the O-DU, and it is considered part of the Midhaul transport architecture by O-RAN WG5. In D-RAN and RAN architectures that do not employ an HLS (Higher Layer Split), then these interfaces will typically be considered part of the Backhaul network.

**Prompt:** Please explain the concept of virtualization in the context of ORAN.

**Generated Response:** In the context of O-RAN (Open Radio Access Network), virtualization refers to the process of running network functions (such as O-RU and O-DU) on software rather than dedicated hardware. This allows for increased flexibility, scalability, and cost-effectiveness by enabling the sharing of resources among multiple network functions and the ability to easily update or replace network functions without affecting the overall system. Virtualization in O-RAN is a key enabler for cloud-native network deployments and enables the realization of the vision of a disaggregated, open, and interoperable RAN.

**Prompt:** Give a step-by-step guide to implementing the O-RAN RIC (RAN Intelligent Controller) API.

**Generated Response:** To implement the O-RAN RIC (RAN Intelligent Controller) API, follow these steps:

1) Understand the O-RAN architecture and its components, including the Near-RT RIC, Non-RT RIC, O-CU, O-CU-CP, O-CU-UP, and DU (Distributed Unit).
2) Familiarize yourself with the O-RAN RIC API specifications, which define the interfaces and functions for controlling and optimizing RAN elements and resources.
3) Implement the Near-RT RIC API, which enables near-real-time control and optimization of RAN elements and resources via fine-grained data collection and actions over E2 interface.
4) Implement the Non-RT RIC API, which enables non-real-time control and optimization of RAN elements and resources, AI/ML workflow including model training and updates, and policy-based guidance of applications/features in Near-RT RIC.
5) Integrate the Near-RT RIC and Non-RT RIC APIs with the O-CU-CP and O-CU-UP components to enable seamless communication and control of the RAN.
6) Test the implementation to ensure that it meets the specified requirements and functions correctly.
7) Deploy the O-RAN RIC API in a network environment and monitor its performance to identify any issues and make necessary adjustments.